
\documentclass[aps,twocolumn,superscriptaddress]{revtex4-1}

\usepackage[T1]{fontenc}

\usepackage{color,xcolor}
\usepackage{soul}

\usepackage{amsmath}

\usepackage{mathrsfs}

\usepackage{physics}
\usepackage{amssymb}
\usepackage{graphicx}

\usepackage[colorlinks=true,linkcolor=blue,citecolor=blue,urlcolor=blue]{hyperref}

\newcommand{\angstrom}{\mbox{\normalfont\AA}}
 
\usepackage{multirow}
\usepackage{float}

\usepackage{appendix}

\makeatletter

\begin{document}
\title{Computational study of samarium hydride at megabar pressures}

\author{Zelong Zhao}
\affiliation{Theory and Simulation of Condensed Matter(TSCM), King's College London, The Strand,London WC2R 2LS,UK}
\author{Siyu Chen}
\affiliation{TCM Group, Cavendish Laboratory, University of Cambridge, J. J. Thomson Avenue, Cambridge CB3 0HE, United Kingdom}
\author{Bartomeu Monserrat}
\affiliation{TCM Group, Cavendish Laboratory, University of Cambridge, J. J. Thomson Avenue, Cambridge CB3 0HE, United Kingdom}
\affiliation{Department of Materials Science and Metallurgy, University of Cambridge, 27 Charles Babbage Road, Cambridge CB3 0FS, United Kingdom}
\author{Evgeny Plekhanov}
\email{evgeny.plekhanov@kcl.ac.uk}
\affiliation{Theory and Simulation of Condensed Matter(TSCM), King's College London, The Strand,London WC2R 2LS,UK}
\author{Cedric Weber}
\email{cedric.weber@kcl.ac.uk}
\affiliation{Theory and Simulation of Condensed Matter(TSCM), King's College London, The Strand,London WC2R 2LS,UK}

\begin{abstract}
Samarium hydride, belonging to the broad class of lanthanide hydrides, has yet to be experimentally tested at high pressure. In this study, we use random structure searches to explore possible stable stoichiometries and propose SmH$_2$ with a layered hexagonal structure in the $P6/mmm$ space group and SmH$_6$ with a hydrogen clathrate structure in the $Im\overline{3}m$ space group as theoretically stable phases of samarium hydride at a wide range of pressures centered around $200$\,GPa. We combine the first principles methods of density functional theory and dynamical mean-field theory to explore many-body correlations in samarium hydride, reporting electron and phonon dispersions and densities of states. Finally, we evaluate the potential electron-phonon driven superconductivity and find low critical temperatures at $200$\,GPa.
\end{abstract}

\maketitle

\section{Introduction}

\par Superconductivity was first observed in the early 20th century at very low temperatures. Scientists have since embarked on a long-standing quest to discover materials with high critical temperatures $T_{\mathrm{c}}$ for practical use. Two mechanisms have been suggested to account for superconductivity: (i) the early proposal of Bardeen-Cooper-Schrieffer\,\cite{bardeen1957theory} in which electrons pair via phonon-mediated interactions, and (ii) spin fluctuations, which account for so-called high-$T_{\mathrm{c}}$ materials such as the copper oxides\,\cite{ginzburg2009theory}.

\par Ashcroft suggested that phonon-mediated superconductors could achieve a high $T_{\mathrm{c}}$ provided that the electron-phonon coupling strength was large~\cite{ashcroft1968metallic}. The initial suggestion was to use high pressure metallic hydrogen~\cite{ashcroft1968metallic}, predicted to have a $T_{\mathrm{c}}$ higher than room temperature \cite{mcmahon2011high}. However, the pressure required to metallize hydrogen exceeds $400$\,GPa, which makes experiments extremely challenging~\cite{dias2017observation}. Ashcroft later suggested that combining hydrogen with other elements would provide additional chemical pressure ~\cite{ashcroft2004hydrogen}, resulting in a reduction of the metallization pressure while maintaining favorable phonon and electron-phonon coupling strengths. First principles calculations have successfully predicted a range of high pressure high temperature superconducting hydrides\,\cite{duan2014pressure,liu2017potential,peng2017hydrogen}, and some of these have been experimentally observed at viable pressures and confirmed to be phonon-mediated\,\cite{drozdov2015conventional,somayazulu2019evidence,drozdov2019superconductivity}. 

\par There are a large number of possible candidates for high pressure hydrides, with a wide range of chemical compositions and different stoichiometries, symmetries, and lattice structures. Amongst all these possibilities, lanthanum hydride \cite{somayazulu2019evidence} has become one of the most studied compounds due to its high $T_{\mathrm{c}}$ of about $250$\,K at $170$\,GPa. The discovery of high temperature superconductivity in lanthanum hydride has motivated studies of many other lanthanide hydrides \cite{fukai2006metal}, both experimentally \cite{somayazulu2019evidence} and computationally \cite{plekhanov2022computational,peng2017hydrogen}. Despite these advances, a particular lanthanide hydride, namely samarium hydride, has so far received little attention. Experimentally, samarium hydride has only been studied at ambient pressure \cite{daou1989low}, while computational studies of samarium hydride under high pressure are still lacking.

\par Crystal structure prediction methods\cite{glass2006uspex,wang2012calypso,pickard2011} have been tremendously successful at identifying the stable phases of high pressure compounds. First principles calculations, typically using density functional theory (DFT), are performed across a wide range of candidate structures and compositions, and the best candidate materials are those with the lowest enthalpy. 

In this study, we use the \textit{ab initio} random structure search ({\sc AIRSS}) methodology~\cite{pickard2006,pickard2011} together with DFT as implemented in the plane wave code {\sc CASTEP} \cite{clark2005first} to predict candidate structures for samarium-hydrogen compounds. The accuracy of the enthalpy determined for each generated structure depends on the quantum solver used, and we use DFT$+U$ \cite{cococcioni2005linear} to describe the local correlations in Sm. We search across the stoichiometry range Sm$_y$H$_x$, with $x=1-18$ and $y=1-2$ and under external pressures ranging from $1$ to $400$\,GPa, as summarized in Fig.\,\ref{fig:database-verification}. Focusing on the $200$\,GPa results, we identify SmH$_2$ in the $P6/mmm$ space group and SmH$_6$ in the $Im\overline{3}m$ space group as stable structures, and study their electron and phonon bands and explore their potential superconductivity. We find that SmH$_2$ has a very low superconducting critical temperature $T_{\mathrm{c}}<1$\,K and SmH$_6$ has a relatively higher $T_{\mathrm{c}}$ above $15$\,K. We also explore the role of electronic correlation in Sm hydrides using DFT augmented with Hubbard $U$ and with dynamical mean field theory, finding correlation effects split the $f$-orbital subspace and reduce the electronic density of states compared to standard DFT. 

\section{Methodology}

\subsection{Theoretical background}

\subsubsection{Electronic structure methods}

\par Within DFT \cite{kohn1965self,hohenberg1964inhomogeneous} the many-body Schr\"{o}dinger equation is solved by mapping the problem onto an auxiliary one-body problem with the same electronic density as the many-body system. This auxiliary Kohn-Sham (KS) system obeys:
\begin{equation}
   \left[-\frac{\hbar^{2}}{2 m} \nabla^{2}+ V_{\mathrm{eff}}(\mathbf{r})\right] \varphi_{i}(\mathbf{r})=\varepsilon_{i} \varphi_{i}(\mathbf{r}), \label{eq:KS} 
\end{equation}
where the electrons experience an effective potential $V_{\mathrm{eff}}(\mathbf{r})$: 
\begin{equation}  
  V_{\mathrm{eff}}(\mathbf{r})=V_{\mathrm{ext}}(\mathbf{r})+\int \frac{n\left(\mathbf{r}^{\prime}\right)}{\left|\mathbf{r}-\mathbf{r}^{\prime}\right|} \mathrm{d}^{3} \mathbf{r}^{\prime}+V_{\mathrm{XC}}[n(\mathbf{r})].
\end{equation}
In these equations, $\varphi_{i}(\mathbf{r})$ is the $i$-th KS orbital, and the electronic density $n(\mathbf{r})$ is derived from the densities of the KS orbitals occupied up to the Fermi level. The accuracy of the KS scheme depends on the choice of exchange correlation (XC) functional. Commonly used approximations include the Local Density Approximation (LDA) and Generalized Gradient Approximations (GGA). In this work, we adopt the Perdew-Burke-Ernzerhof (PBE) XC functional \cite{perdew1996generalized}, which is a type of GGA.

\par The amount of correlations contained in the DFT XC functional is not sufficient to treat the strong Coulomb repulsion in partially filled $d$ and $f$ orbitals. Within standard approximations to DFT functionals, these strongly correlated orbitals appear to be excessively delocalized. This lack of localization can be, to some extent, corrected with the DFT+U scheme, where the DFT energy functional is combined with an additional term proportional to a parameter, called Hubbard $U$, which  penalizes the configurations with doubly occupied orbitals~\cite{cococcioni2005linear}. 

\par A higher level treatment of the strong Coulomb repulsion is provided by the DFT + Dynamical Mean Field Theory (DFT+DMFT) approach~\cite{georges1996dynamical,kotliar2006electronic,evgeny2018}, where the temporal correlations are taken into account exactly, while the treatment of the spatial correlations becomes exact in the limit of infinite coordination of the correlated orbitals. Within DFT+DMFT, the system of correlated orbitals (usually $d$ or $f$, or a subset of them) connected to a chemical environment is mapped onto an auxiliary problem of an Anderson impurity connected to a bath of uncorrelated orbitals. The latter problem is then solved either numerically (e.g. quantum Monte Carlo, exact diagonalization) or analytically by using some approximation (e.g. Hubbard-I, auxiliary bosons). The results of such an impurity problem are then mapped back onto the original correlated lattice problem. The details of the DFT+DMFT implementation used in the present work can be found in Refs.~\cite{evgeny2018,evgeny_forces}.

\subsubsection{Structure searching}

\par We perform structure searches using {\sc AIRSS}~\cite{pickard2006,pickard2011} together with the DFT+U methodology. The efficiency of the structure search is crucial for searching low enthalpy compounds, and we adopt several strategies to accelerate calculations. These include setting constraints on the maximum unit cell volume and minimum bond length when generating random structures. Additionally, we utilized low-resolution calculations during the initial potential energy surface scan, which were then followed by higher-precision calculations. To obtain the convex hull of samarium hydride, we calculate the enthalpy of formation with:

\begin{equation}\label{eq:formation_enthalpy}
	\Delta H = H_{\text{Sm}_y\text{H}_x} -x \, H_\text{H} - y \, H_\text{Sm}.
\end{equation}
This formula requires the knowledge of stable phases of the end members samarium \cite{finnegan2020high} and hydrogen \cite{pickard2007structure}, which we take from the literature. For example, at $200$\,GPa, hydrogen is in a phase of $C2/c$ symmetry and samarium is in the $oF8$ phase.

\par  We highlight that it is generally impossible to confirm whether the global enthalpy minimum has been found due to the exponential scaling cost of searching over the potential energy landscape. However, over the past decade structure searching methods have been shown to provide important insights into high pressure phases and often predicted the correct structures subsequently identified experimentally.

\subsection{Computational details}

\par In the present work, we perform the first principles calculations using the {\sc CASTEP} and {\sc Quantum Espresso} codes. We use the Perdew-Burke-Ernzerhof (PBE) exchange-correlation function together with ultrasoft pseudopotentials for Sm [Xe]4f$^6$6s$^2$ and H 1s$^1$.

\par We perform a random structure search over the samarium-hydrogen binary combining {\sc AIRSS} \cite{pickard2011} and {\sc CASTEP} \cite{clark2005first}. We adopt a two-step approach to facilitate the random structure search with DFT+U: (1) we explore different stoichiometries using a coarse Monkhorst-Pack (MP) \textbf{k}-point grid (2$\pi \times$ 0.03 \AA$^{-1}$), a kinetic energy cut-off of $E_{\mathrm{cut}}=500$\,eV, a force tolerance of $0.05$\,eV/\AA; and (2) we perform additional calculations on a small subset of structures having the lowest enthalpy with a denser MP \textbf{k}-point grid (2$\pi \times$ 0.01 \AA$^{-1}$), $E_{\mathrm{cut}}=750$\,eV, force tolerance $0.001$\,eV/\AA, and $U=6$\,eV (Fig.\,\ref{fig:high_precision_vs_low}). We note that the parameter set used in the second step ensures that the overall energy precision is within $1$ meV/atom (see Fig. \ref{fig:KE_K_convergence_P6mmm}, Fig. \ref{fig:KE_K_convergence_Im3m}).

\par For the pressure of 200~GPa, we also perform DFT+DMFT calculations as implemented within {\sc CASTEP}~\cite{clark2005first,evgeny2018,evgeny_forces} using the same DFT parameters as the ones used in step (2) of the searches. Additionally, these calculations used a Hubbard $U=6$ eV and an additional Hund's coupling $J=0.855$ eV \cite{banerjee2022pressure}. 

\par We calculate phonon dispersions in the harmonic approximation via the finite difference method~\cite{Parlinski1997} combined with nondiagonal supercells~\cite{Lloyd2015}. The 3-dimensional Farey \textbf{q}-grid of order 6 is used for careful sampling of the dynamical matrices~\cite{chen2022nonuniform} and confirming the dynamical stability of the systems. The electron-phonon coupling properties were computed via density functional perturbation theory (DFPT), which was implemented using the {\sc Quantum Espresso} package\,\cite{giannozzi2009quantum}. The uniform \textbf{q}-point grids used to sample the Brillouin zone are of size $6\times6\times6$. The DFT settings involved in the DFPT calculation were identical to those used in step (2) of the searches. The superconducting critical temperature was found using the Allen and Dynes~\cite{allen1975transition} revised McMillan~\cite{mcmillan1968transition} equation.

\section{Result and discussion}

\subsection{Structure Search}

\par
\begin{figure}[h!]
    \centering
\includegraphics[width=1.\columnwidth]{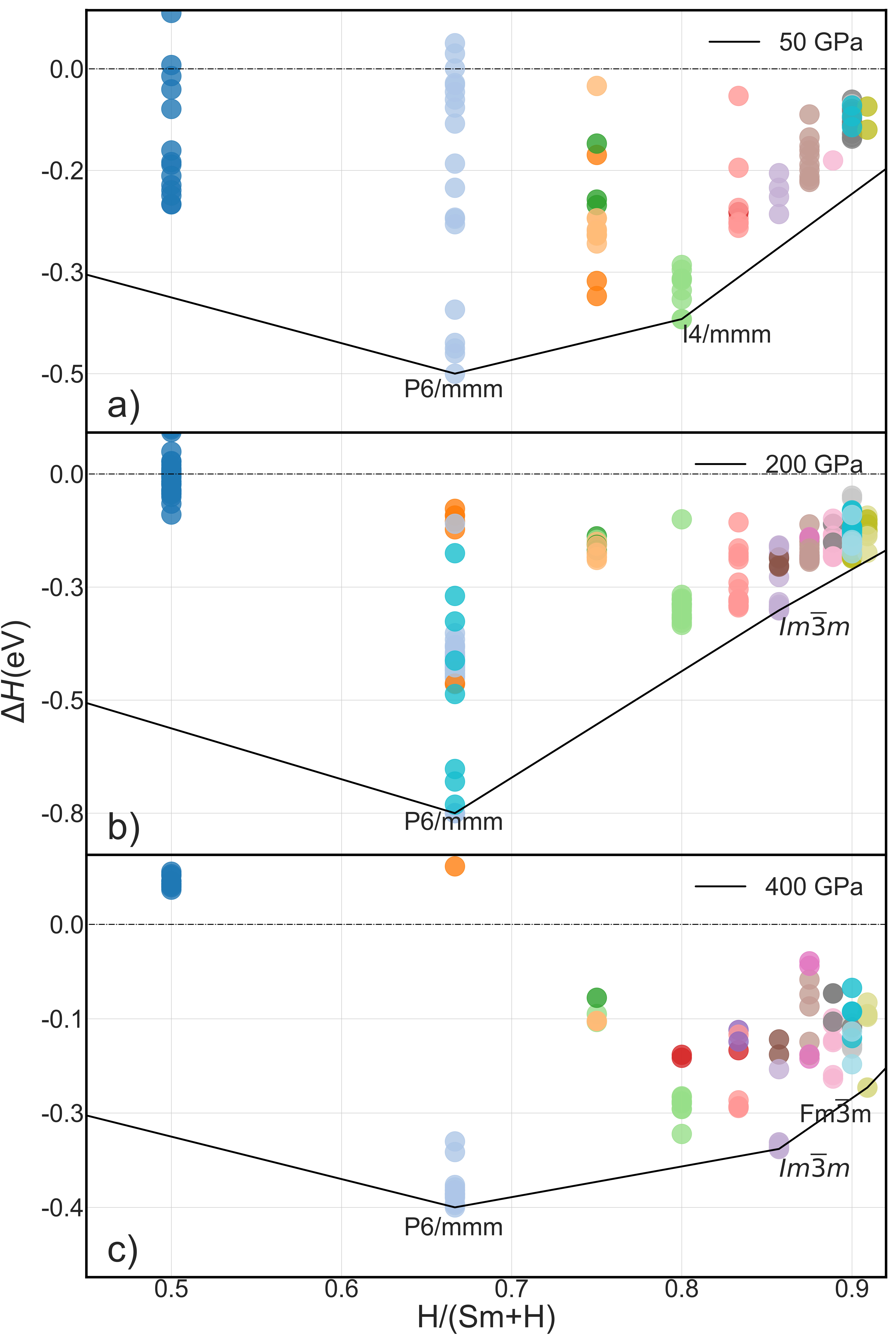}
    \caption{Maxwell convex hull reported at $50$, $200$, and $400$\,GPa. The solid line connects stable phases and the dashed line connects the lowest enthalpy phases at each stoichiometry (not necessarily stable). The color of each circle indicate the space group of the phase. The end members are hydrogen in the  $C2/c$ structure \cite{pickard2007structure} and samarium in the $of8$ phase \cite{finnegan2020high}. }
    \label{fig:convexhull}
\end{figure}

 The results of the structure searches are summarized by the convex Hull diagrams shown in Fig.\,\ref{fig:convexhull}. Each point corresponds to a distinct structure at a certain pressure and stoichiometry. In the convex Hull, the horizontal axis is the ratio $x$ of hydrogen atoms relative to the total number of atoms in the cell in one formula unit (f.u.). The left side, corresponding to $x=0$, is the samarium bulk, and the right side, corresponding to $x=1$, is the hydrogen bulk (note that the pure phases are outside the range depicted in Fig.\,\ref{fig:convexhull}). The solid line connects thermodynamically stable phases.

%Fm-3m 400
% table here wrote each phase at which pressure were previous found stable and maybe Tc as well indicate experinment work or theoretical prediction. 
% I4/mmm high than 170 GPa 100k Tc computational
% YH$_6$ Im-3m 180 GPa 220k experiment 
% LaH10 \cite{sun2021high (experinment) Fm-3m higher than 138 GPa

\par We performed structure searches at three different pressures of $50$, $200$, and $400$\,GPa (Figure. \ref{fig:convexhull}). We remark that the enthalpies of elemental hydrogen and samarium enter the calculation of the formation energy, and we use the theoretical estimates of solid hydrogen from Ref.\,\cite{pickard2007structure} and the $oF8$ structure of samarium reported in Ref.\,\cite{finnegan2020high}. We find that SmH$_2$ in a $P6/mmm$ structure appears consistently as a stable phase throughout the tested pressure range. At a pressure of $50$\,GPa we also find SmH$_4$ in a $I4/mmm$ structure as a stable phase. At a pressure of $200$\,GPa, SmH$_4$ is no longer stable, and instead a SmH$_6$ structure in the $Im\overline{3}m$ space group becomes stable. At the highest pressure studied of $400$\,GPa, SmH$_2$ and SmH$_6$ are still stable, and another SmH$_{10}$ phase in the $Fm\overline{3}m$ space group also becomes stable. As pressure levels rise, we observe stable compounds with increasing hydrogen content \cite{ashcroft1968metallic}.

\par The stoichiometries and space groups of the stable phases we have discovered in our searches have also been observed or predicted in other high-pressure lanthanide hydride systems. The SmH$_2$ phase with $P6/mmm$ space group parallels the ScH$_2$ phase described in Ref.,\cite{peng2017hydrogen}. The SmH$_4$ phase with $I4/mmm$ space group has also been predicted as a stable phase in ScH$_4$ at pressures exceeding $170$\,GPa\,\cite{qian2017theoretical}. The SmH$_6$ phase with $Im\overline{3}m$ space group parallels an yttrium hydride phase observed experimentally around $200$\,GPa\,\cite{kong2021superconductivity}. Finally, a LaH$_{10}$ phase in space group $Fm\overline{3}m$ has also been observed experimentally \cite{sun2021high}. 

\par We also note that the Sm-H convex Hull was previously reported in Ref.\,\cite{peng2017hydrogen}. However, this earlier report was not based on a structure search of the Sm-H system; instead structure searches were performed for other lanthanide-hydrogen systems, and the lanthanide elements were subsequently replaced by Sm in the final structures. Of the Sm-H structures reported in Ref.\,\cite{peng2017hydrogen}, we have re-discovered the SmH$_4$ $I4/mmm$, SmH$_6$ $Im\overline{3}m$, and SmH$_{10}$ $Fm\overline{3}m$  phases in our searches, and identified for the first time the stable SmH$_2$ $P6/mmm$ phase. Our results are also consistent with the observation in Ref.\,\cite{peng2017hydrogen} that hydrogen-rich phases become competitive at higher pressures.

\begin{figure}[h!]
    \centering
\includegraphics[width=1.\columnwidth]{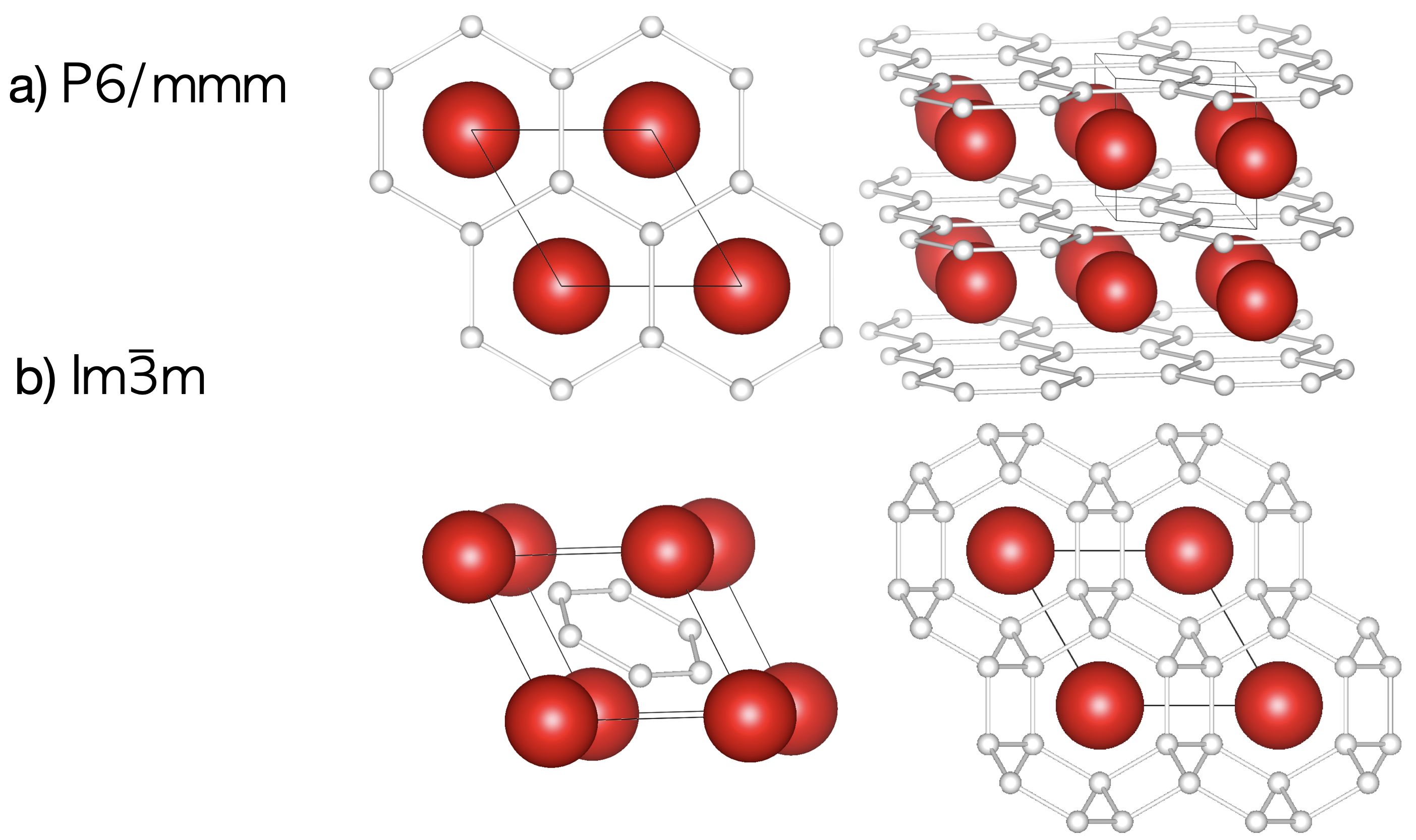}
    \caption{(a) Crystal structure of SmH$_2$ in space group $P6/mmm$ and (b) crystal structure of SmH$_6$ in space group $Im\overline{3}m$ at $200$\,GPa.}
    \label{fig:fig2}
\end{figure}

In the rest of this work, we focus our analysis on the pressure of $200$\,GPa, a pressure range commonly employed in high-pressure hydride experiments. Spin-Orbital Coupling (SOC)~{\cite{steiner2016calculation}} has been identified as an important factor in heavy fermionic systems, particularly when studying the excitation properties of samarium chalcogenides~{\cite{kang2016electronic}}. While its role is crucial in these contexts, our focus is on the study of structural properties and formation enthalpies. For such specific studies, SOC can be safely neglected, as the SOC energy contribution several orders of magnitude smaller than Coulomb interaction in the samarium $4f$ electrons of samarium chalcogenides~{\cite{banerjee2022pressure}}. Furthermore, in our implementation of the DMFT solver, we capture both  $J=7/2$ and $J=5/2$ multiplets in a larger correlated subspace without explicitly identifying them. Given these considerations, our attention is focused on the two prominent phases at 200 GPa: SmH$_2$ in the space group $P6/mmm$ and SmH$_6$ in the space group $Im\overline{3}m$, both of which are shown in Fig.\,\ref{fig:fig2}.

\subsection{Electronic properties of SmH$_2$ $P6/mmm$ and SmH$_6$ $Im\overline{3}m$}

\par
\begin{figure}[h]
    \centering
\includegraphics[width=1.\columnwidth]{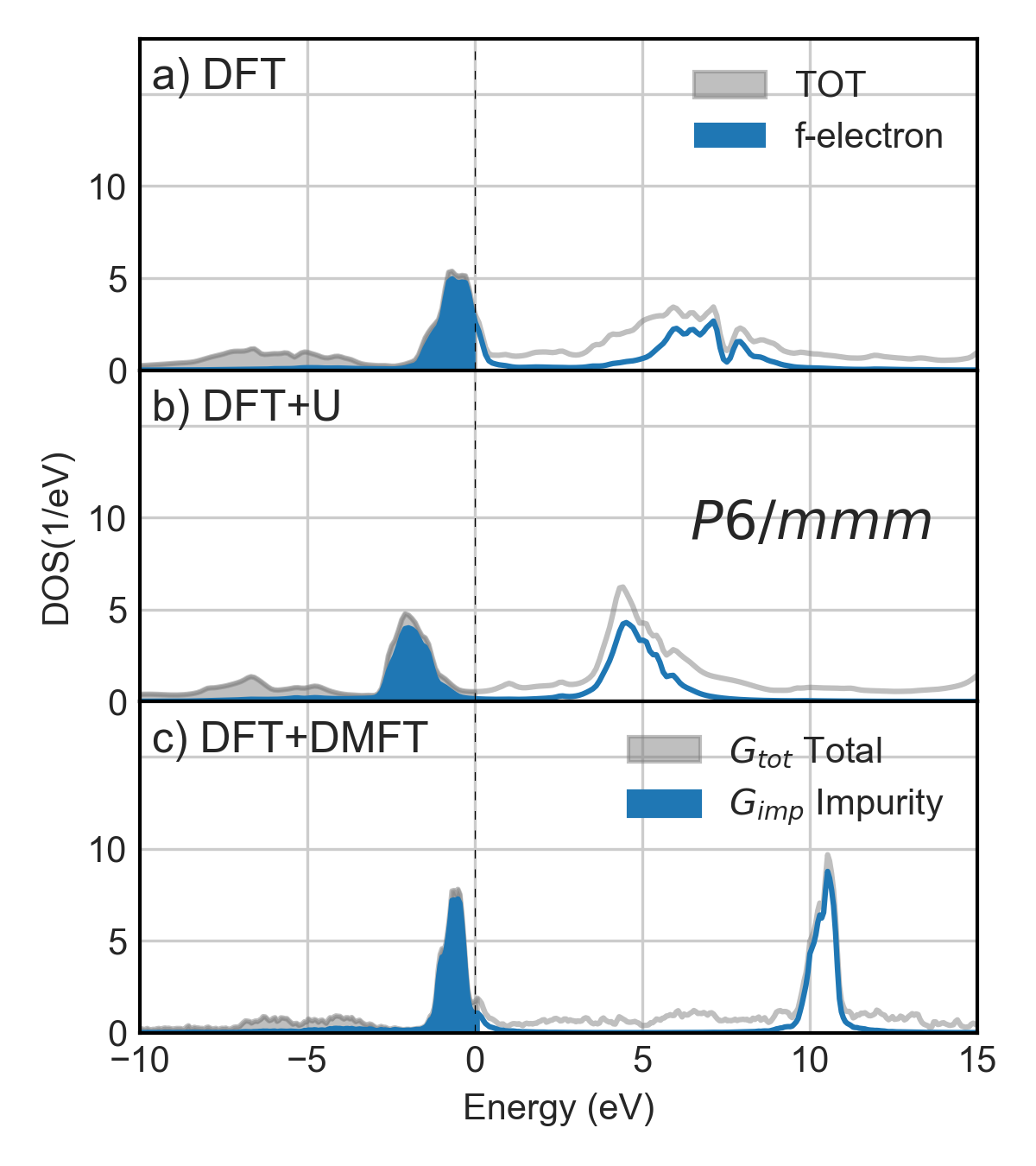}
    \caption{Density of states of SmH$_2$ $P6/mmm$ at 200\,GPa as predicted by (a) DFT and (b) DFT+U with $U=6$\,eV. TOT labels the total DOS and $f$-electron labels the partial DOS from samarium $f$ electrons. (c) Spectral function calculated using DFT+DMFT, incorporating a Hubbard $U$ of 6\,eV and Hund's coupling $J=0.85$\,eV. G$_{\mathrm{tot}}$ represents the spectral function derived from the total Green's function, while G$_{\mathrm{imp}}$ indicates the spectral function obtained from the impurity Green's function.}
    \label{fig:P6mmm_dos}
\end{figure}

\begin{figure}[h]
    \centering
\includegraphics[width=1.\columnwidth]{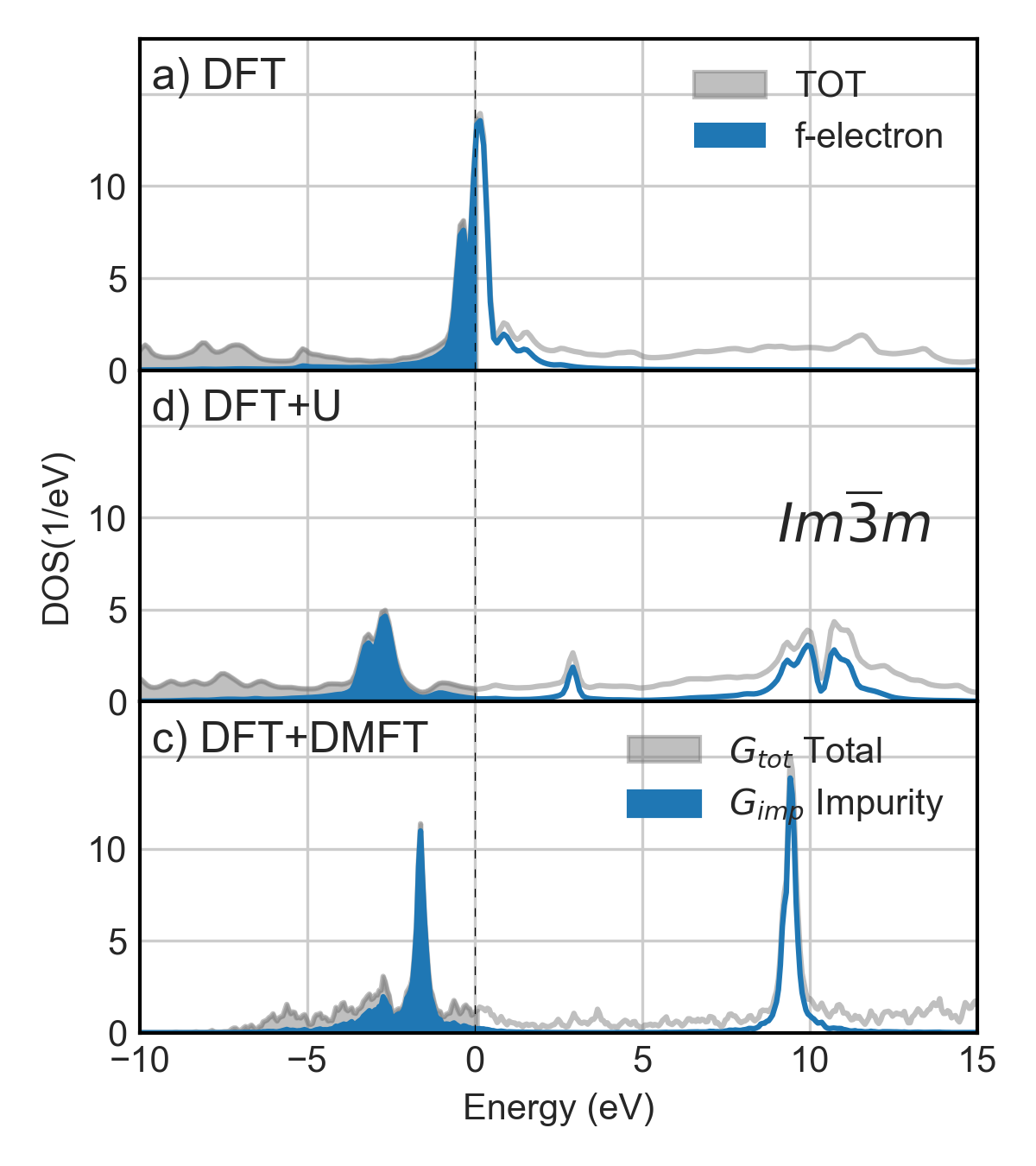}
    \caption{Density of states of SmH$_6$ $Im\overline{3}m$ at 200\,GPa as predicted by (a) DFT and (b) DFT+U with $U=6$\,eV. TOT labels the total DOS and $f$-electron labels the partial DOS from samarium $f$ electrons. (c) Spectral function calculated using DFT+DMFT, incorporating a Hubbard $U$ of 6\,eV and Hund's coupling $J=0.85$\,eV. G$_{\mathrm{tot}}$ represents the spectral function derived from the total Green's function, while G$_{\mathrm{imp}}$ indicates the spectral function obtained from the impurity Green's function.}
    \label{fig:Im3m_dos}
\end{figure}

 To investigate the electronic properties of the SmH$_2$ $P6/mmm$ and SmH$_6$ $Im\overline{3}m$ phases, we calculate their densities of states (DOS) and show them in Figs.\,\ref{fig:P6mmm_dos} and \ref{fig:Im3m_dos}. DFT calculations using PBE do not treat the potential correlation of f-orbital electrons, and as a result the electronic structures of $P6/mmm$ (Fig.\,\ref{fig:P6mmm_dos}a) and $Im\overline{3}m$ (Fig.\,\ref{fig:Im3m_dos}a) show a peak at the Fermi level dominated by $f$-orbital electrons. The Hubbard U term in the DFT+U method drives a Mott transition, splitting the DOS of the correlated orbitals and pushing the associated peaks in the density of states away from the Fermi level. 

\par To further explore correlation within the $f$ electron orbitals, we perform DFT+DMFT calculations for SmH$_2$ $P6/mmm$ and SmH$_6$ $Im\overline{3}m$ at $200$\,GPa. We employ charge self-consistent (CSC) DFT+DMFT, which involves the convergence between the chemical potential in the DMFT calculation of the impurity and the charge density obtained from the self-consistent field calculation of the bath. Our calculations show qualitatively similar results between DFT+U and DFT+DMFT, with a Mott-like gap opening in both cases. Quantitatively, the gap is larger in the DFT+DMFT case, but the electronic density of states near the Fermi level is comparable. These results suggest that DFT+U and DFT+DMFT should lead to similar results for the electronic properties of samarium hydride compounds. 
%show similar densities of states near the Fermi level. It's important to acknowledge that similar pattern on DOS at the Fermi level doesn't necessarily produce identical results for other properties such as (T$_{\mathrm{c}}$) \cite{plekhanov2022computational}.

%Firstly explain how DFT+U DOS are found and DFT+DMFT how to it works. When the two density of state are similiar what is that mean. 
%1 DFT+U DOS
%2 DFT+DMFT DOS
%3 What is that MEAN

%From the comparison of DFT and DMFT DOSes, you can immediately see that there is opening of a Mott insulating gap in the latter case, because DFT is known to miss the proper strong correlations treatment. This indicated that 

\begin{figure}[h]
    \centering
\includegraphics[width=1.\columnwidth]{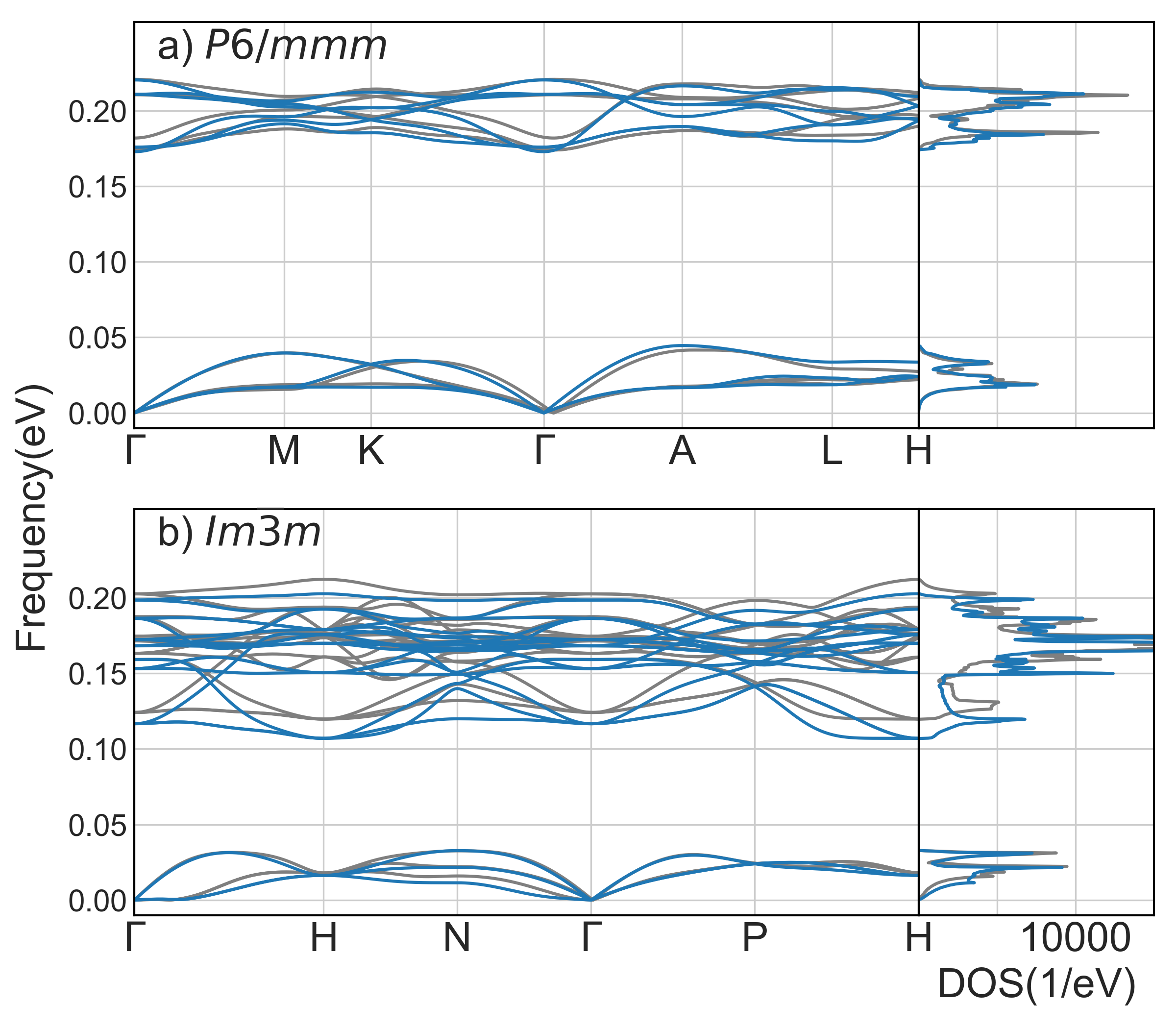}
    \caption{Phonon dispersion (left panels) and density of states (right panels) for $P6/mmm$ SmH$_2$ and $Im\overline{3}m$ SmH$_6$, calculated using DFT and DFT+$U$ for values of $U=6$\,eV. Gray and blue curves represent DFT and DFT+$U$ respectively.}
    \label{fig:fig_phono}
\end{figure}

\subsection{Vibrational properties of SmH$_2$ $P6/mmm$ and SmH$_6$ $Im\overline{3}m$}

\par The phonon dispersions of SmH$_2$ $P6/mmm$ and SmH$_6$ $Im\overline{3}m$ at $200$\,GPa using DFT and DFT+$U$ are shown in Fig.\,\ref{fig:fig_phono}. The dispersions exhibit no soft modes, which indicates that the predicted phases are dynamically stable. The phonons of samarium hydride show a similar pattern to those of other lanthanides hydrides. The most prominent feature is a large gap between optical and acoustic branches arising from the large difference in mass between lanthanides and hydrogen. Interestingly, the Hubbard $U$ term has a negligible effect on the phonon dispersion. We also highlight that different values of $U$ lead to an equilibrium volume for the $P6/mmm$ structure that varies from $15.72\,\angstrom^3$ to $16.09\,\angstrom^3$ for $U$ between $0$\,eV and $6$\,eV and for the $Im\overline{3}m$ structure that varies from $21.35\,\angstrom^3$ to $22.45\,\angstrom^3$. Given the effect of including Hubbard U functional on volume, the decrease in phonon frequency, as shown in Fig.{~\ref{fig:fig_phono}}, is consistent with volume contraction.

\subsection{Superconductivity of SmH$_2$ $P6/mmm$ and SmH$_6$ $Im\overline{3}m$}

\par We perform density functional perturbation theory (DFPT) with DFT calculations to estimate the electron-phonon coupling strength and calculate the superconducting temperature ($T_{\mathrm{c}}$) via the Allen and Dynes \cite{allen1975transition} revised McMillan \cite{mcmillan1968transition} equation: 

\begin{equation}\label{eq:mcmillan}
T_{\mathrm{c}}=\frac{\omega_{\mathrm{log}}}{1.2} \exp \left[\frac{-1.04(1+\lambda)}{\lambda\left(1-0.62 \mu^{*}\right)-\mu^{*}}\right]
\end{equation}
In this expression, $\mu^{*}$ is the Coulomb potential with typical values between $0.1$ and $0.15$, $\lambda$ is the electron-phonon coupling constant, and $\omega_{\mathrm{log}}$ is the logarithmic average frequency. These quantities are evaluated via integrals of the Eliashberg function.

\begin{table}[]
\centering
\caption{Electron-phonon coupling of Samarium hydrides predicted by DFT+DFPT.}
\label{tab:tc}
\begin{tabular}{cccc}
\hline
\hline
 & $\lambda$ & $\omega_{log}$ & $T_c(K)$ \\ \hline
\multirow{2}{*}{$P6/mmm$} & \multirow{2}{*}{0.28} & \multirow{2}{*}{357.69} & 0.10 ($K$) ($\mu^{*}=0.10$) \\
 &  &  & 0.00 ($K$) ($\mu^{*}=0.15$) \\ \hline
\multirow{2}{*}{$Im\overline{3}m$} & \multirow{2}{*}{0.65} & \multirow{2}{*}{900.04} & 26.35 ($K$) ($\mu^{*}=0.10$) \\
 &  &  & 15.50 ($K$) ($\mu^{*}=0.15$) \\ \hline\hline

\end{tabular}
\end{table}

\par In $P6/mmm$ SmH$_2$, the electron-phonon coupling constant is relatively low, with $\lambda=0.28$, resulting in a $T_{\mathrm{c}}$ value below $1$\,K for $\mu^{*}$ ranging between $0.1$ and $0.15$. The $Im\overline{3}m$ SmH$_6$ structure exhibits a higher electronic density of states at the Fermi level, leading to a comparatively higher predicted $T_{\mathrm{c}}$ between $15$ and $26$\,K depending on the value of $\mu^{*}$. The relatively low superconducting critical temperatures in the Sm-H system are consistent with earlier work \cite{peng2017hydrogen,semenok2020distribution} which shows higher $T_{\mathrm{c}}$ for La/Y hydrides and lower $T_{\mathrm{c}}$ for intermediate series lanthanides. The origin of the low superconducting critical temperature in samarium hydride could be caused by the low hydrogen DOS at the Fermi level and the associated weak electron-phonon coupling strength. 

\section{Conclusion}

\par In conclusion, we have performed random structure searches to explore potential phases of the high pressure samarium-hydrogen system. Focusing on a pressure of $200$\,GPa, we predict that SmH$_2$ with a structure of space group $P6/mmm$ and SmH$_6$ with a structure of space group $Im\overline{3}m$ are the stable phases. We also find other stable phases of stoichiometries SmH$_4$ and SmH$_{10}$ at different pressures. We have conducted a comprehensive analysis of the electronic structure and lattice dynamics of SmH$_2$ and SmH$_6$ at $200$\,GPa. Our finding reveals that electronic correlation plays an important role in splitting the f-electron density of state. Indeed, for the lanthanide mid-series (nearly half-shell), we expect a large collection of low-energy electronic states associated with the paramagnetic multiplets, leading to paramagnetism at room temperature and high pressure. The latter are inherently multi-determinantal states not well captured by DFT or DFT+U, but well treated within the non-perturbative DMFT approximation. A fingerprint or such physics is the splitting of the f-bands into sub-structures, some of which pushed closer or further away from the Fermi level and in turn inducing large corrections in the calculations of the superconducting temperature. In our work, our calculations however validate a low superconducting temperature as obtained by DFT. The superconducting critical temperature of SmH$_2$ is below $1$\,K, while for SmH$_6$ it falls within the range $15$ to $26$\,K.

\par This work opens new avenues for studying hydrides at high pressure. Specifically, we highlight the importance that electronic correlation can play in the electronic properties of compounds with partially filled $f$ electron shells. The method used in this work is general and can be extended to other systems of interest.

\quad

\section*{Acknowledgements}
C.W. and E.P. are supported by the grant [EP/R02992X/1] from the UK Engineering and Physical Sciences Research Council (EPSRC). S.C. acknowledges financial support from the Cambridge Trust and from the Winton Programme for the Physics of Sustainability. B.M. acknowledges support from a UKRI Future Leaders Fellowship (Grant No. MR/V023926/1), from the Gianna Angelopoulos Programme for Science, Technology, and Innovation, and from the Winton Programme for the Physics of Sustainability. This work was performed using resources provided by the ARCHER UK National Supercomputing Service and the Cambridge Service for Data Driven Discovery (CSD3) operated by the University of Cambridge Research Computing Service (www.csd3.cam.ac.uk), provided by Dell EMC and Intel using Tier-2 funding from the Engineering and Physical Sciences Research Council (capital grant EP/P020259/1), and DiRAC funding from the Science and Technology Facilities Council (www.dirac.ac.uk).

% \bibliography{main-bib}
%

\clearpage

%%%%%%%%%% Merge with supplemental materials %%%%%%%%%%
\widetext
\begin{center}
\textbf{\large Supplemental Materials: Computational prediction of Samarium Hydride at Magebar pressure}
\end{center}

\setcounter{equation}{0}
\setcounter{figure}{0}
\setcounter{table}{0}
\setcounter{page}{1}
\setcounter{section}{0}
\makeatletter
\renewcommand{\theequation}{A\arabic{equation}}
\renewcommand{\thefigure}{A\arabic{figure}}
\renewcommand{\thetable}{A\arabic{table}}

\section{Random structure search}

\begin{figure}[h!]
    % \centering
\includegraphics[width=0.8\linewidth]{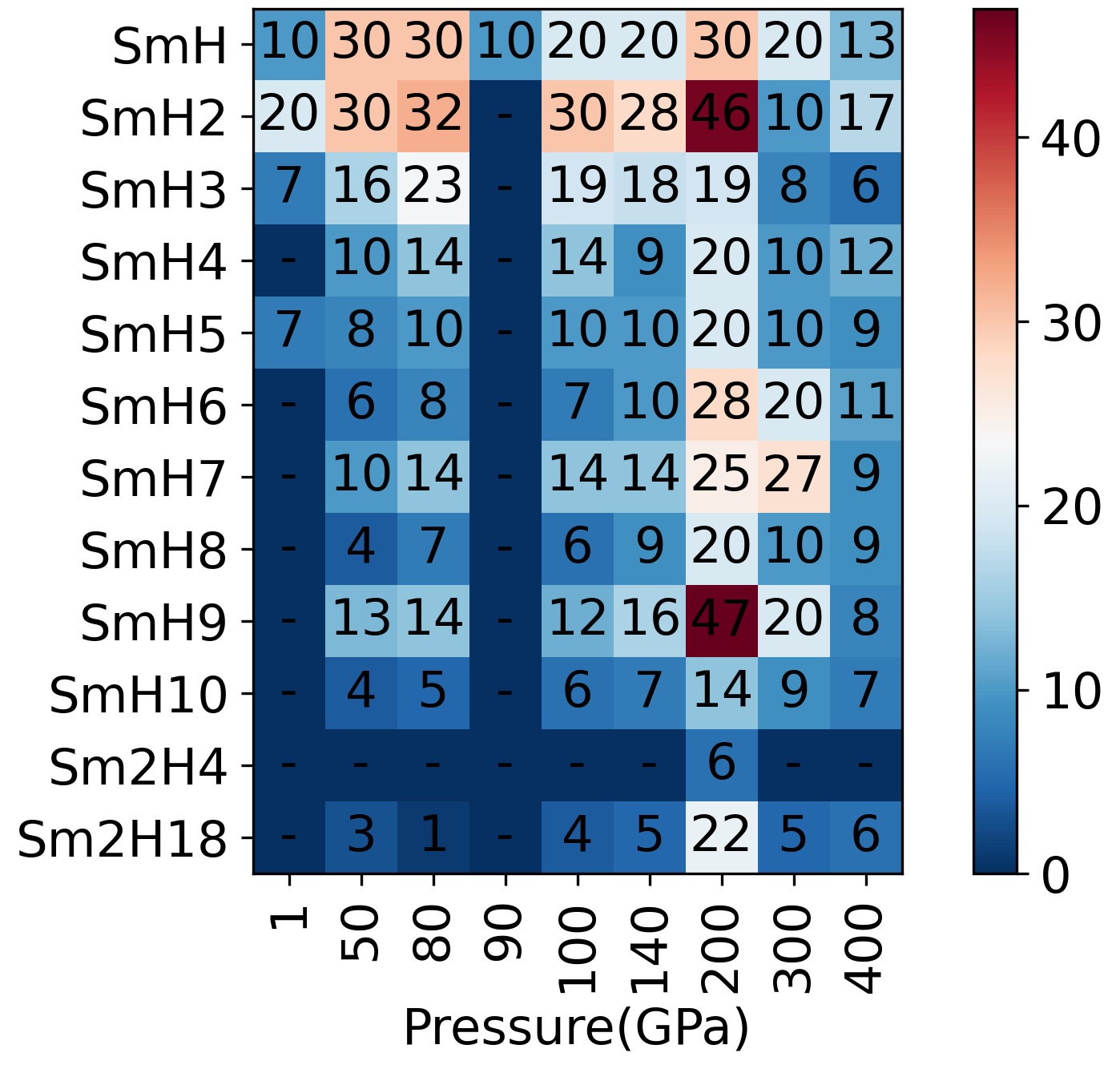}
    \caption{Size of the phase database. Y-axis is stoichiometry of Samarium and hydrogen. X-axis is pressure we performed random structure search. Phase we searched "-" represents we did not search for this pressure and stoichiometry.  The number indicated in each of the cell is the total number of distinct structure identified by AIRSS at each pressure for each stoichiometry. Colour bar blue for less search performed for this combination and red represents for more intense search.}
    \label{fig:database-verification}
\end{figure}

\begin{figure}[h!]
\includegraphics[width=0.8\linewidth]{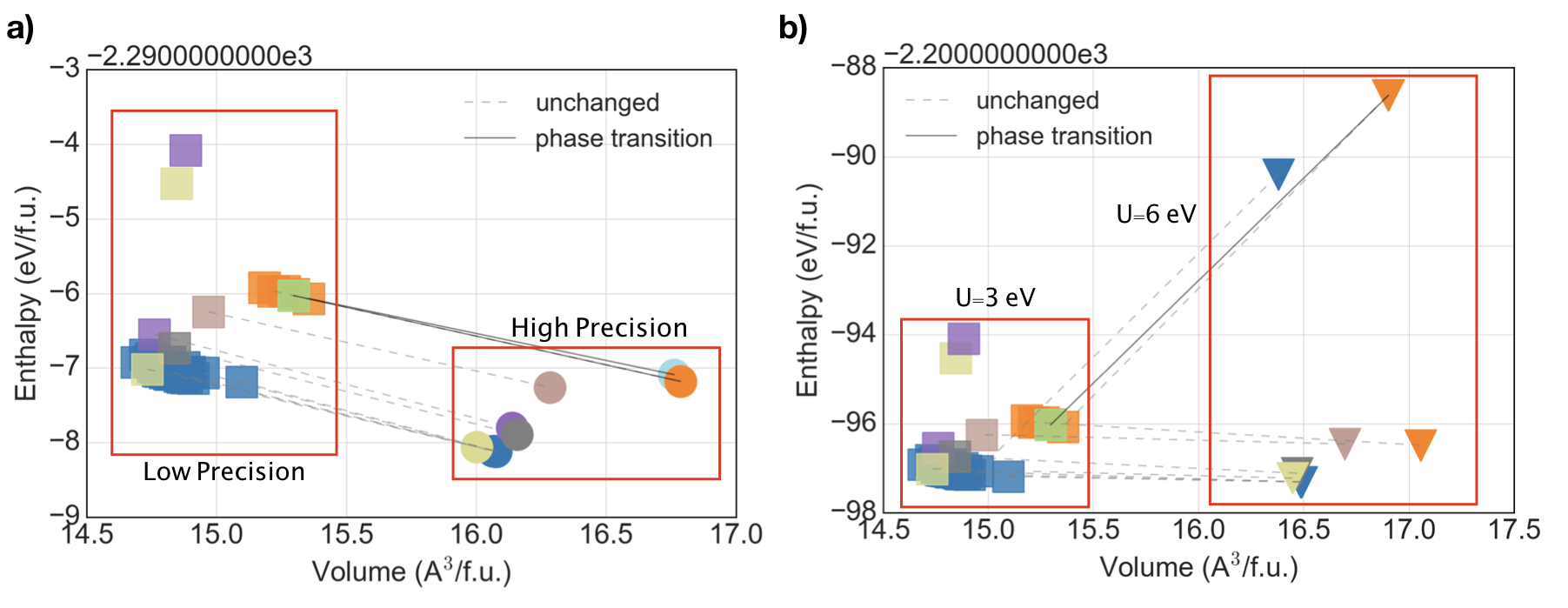}
\caption{AIRSS generated phases of SmH$_2$ at 200\,GPa. Squares and circles indicate phases of SmH$_2$. Color indicates space group. Square indicate relevant DFT+U settings is "step 1" as mentioned in the main paper and circles represent "step 2". X axis is volume of the unit cell and Y axis is DFT+U calculated enthalpy. Phases in both "Low Precision and (b) "U=3eV" boxes are same group of phases and DFT+U settings are indicated are "step 1" in the method section. (a) DFT settings changed between "Low Precision" ("step 1") to "High Precision" ("step 2") but Hubbard $U$ is kept same. (b) DFT settings changed as in (a) but Hubbard $U$ involved in DFT+U calculation changed from 3~eV to 6~eV.}
\label{fig:high_precision_vs_low}
\end{figure}

\begin{table}[h!]
\centering
\caption{Lattice parameters of Sm$_x$H$_y$}
\label{tab:my-table}
\begin{tabular}{ccccccc}
\hline
 & \begin{tabular}[c]{@{}c@{}}Space \\ group\end{tabular} & \begin{tabular}[c]{@{}c@{}}Lattice \\ Parameters \AA\end{tabular} & Atom & \multicolumn{3}{c}{Atomic fractional coordinates} \\
 &  &  &  & X & Y & Z \\ \hline
\begin{tabular}[c]{@{}c@{}}SmH$_2$\\ (200 GPa)\end{tabular} & $P6/mmm$ & \begin{tabular}[c]{@{}c@{}}a=b=2.692\\ c=2.562\\ $\alpha=\beta=90^{\circ}$\\ $\gamma =  120^{\circ}$\end{tabular} & \begin{tabular}[c]{@{}c@{}}Sm(1b)\\ H(2c)\end{tabular} & \begin{tabular}[c]{@{}c@{}}0.000\\ 1/3\end{tabular} & \begin{tabular}[c]{@{}c@{}}0.000\\ 2/3\end{tabular} & \begin{tabular}[c]{@{}c@{}}0.000\\ 1/2\end{tabular} \\ \hline
\begin{tabular}[c]{@{}c@{}}SmH$_4$\\ (200 GPa)\end{tabular} & $I4/mmm$ & \begin{tabular}[c]{@{}c@{}}a=b=2.546\\ c=5.618\\ $\alpha=\beta=\gamma=90^{\circ}$\end{tabular} & \begin{tabular}[c]{@{}c@{}}Sm(2b)\\ H(4e)\\ H(4d)\end{tabular} & \begin{tabular}[c]{@{}c@{}}0.000\\ 1/2\\ 0.000\end{tabular} & \begin{tabular}[c]{@{}c@{}}0.000\\ 1/2\\ 1/2\end{tabular} & \begin{tabular}[c]{@{}c@{}}1/2\\ 0.693\\ 3/4\end{tabular} \\ \hline
\begin{tabular}[c]{@{}c@{}}SmH$_4$\\ (100 GPa)\end{tabular} & $C2/m$ & \begin{tabular}[c]{@{}c@{}}a=5.553\\ b=2.842\\ c=3.041\\ $\alpha=\gamma=90^{\circ}$\\ $\beta=112.492^{\circ}$\end{tabular} & \begin{tabular}[c]{@{}c@{}}Sm(2a)\\ H(4i)\\ H(4i)\end{tabular} & \begin{tabular}[c]{@{}c@{}}1/2\\ 0.901\\ 0.747\end{tabular} & \begin{tabular}[c]{@{}c@{}}1/2\\ 1/2\\ 1/2\end{tabular} & \begin{tabular}[c]{@{}c@{}}0.000\\ 0.455\\ 0.669\end{tabular} \\ \hline
\begin{tabular}[c]{@{}c@{}}SmH$_4$\\ (140 GPa)\end{tabular} & $Fmmm$ & \begin{tabular}[c]{@{}c@{}}a=3.792\\ b=4.141\\ c=5.238\\ $\alpha=\beta=\gamma=90^{\circ}$\end{tabular} & \begin{tabular}[c]{@{}c@{}}Sm(4b)\\ H(8f)\\ H(8i)\end{tabular} & \begin{tabular}[c]{@{}c@{}}0.000\\ 3/4\\ 1/2\end{tabular} & \begin{tabular}[c]{@{}c@{}}0.000\\ 3/4\\ 1/2\end{tabular} & \begin{tabular}[c]{@{}c@{}}1/2\\ 3/4\\ 0.073\end{tabular} \\ \hline
\begin{tabular}[c]{@{}c@{}}SmH$_6$\\ (140 GPa)\end{tabular} & $Im\overline{3}m$ & \begin{tabular}[c]{@{}c@{}}a=b=c=3.603\\ $\alpha=\beta=\gamma=90^{\circ}$\end{tabular} & \begin{tabular}[c]{@{}c@{}}Sm(2a)\\ H(12d)\end{tabular} & \begin{tabular}[c]{@{}c@{}}1/2\\ 1/2\end{tabular} & \begin{tabular}[c]{@{}c@{}}1/2\\ 1/4\end{tabular} & \begin{tabular}[c]{@{}c@{}}1/2\\ 0.000\end{tabular} \\ \hline
\begin{tabular}[c]{@{}c@{}}SmH$_{10}$\\ (400 GPa)\end{tabular} & $Fm\overline{3}m$ & \begin{tabular}[c]{@{}c@{}}a=b=c=4.434\\ $\alpha=\beta=\gamma=90^{\circ}$\end{tabular} & \begin{tabular}[c]{@{}c@{}}Sm(4a)\\ H(8c)\\ H(32f)\end{tabular} & \begin{tabular}[c]{@{}c@{}}1/2\\ 3/4\\ 0.620\end{tabular} & \begin{tabular}[c]{@{}c@{}}0.000\\ 1/4\\ 0.620\end{tabular} & \begin{tabular}[c]{@{}c@{}}1/2\\ 1/4\\ 0.380\end{tabular} \\ \hline
\end{tabular}
\end{table}

\pagebreak

\section{DFT+DMFT Calibration}

\begin{figure}[th!]
    % \centering
    \includegraphics[width=1.\linewidth]{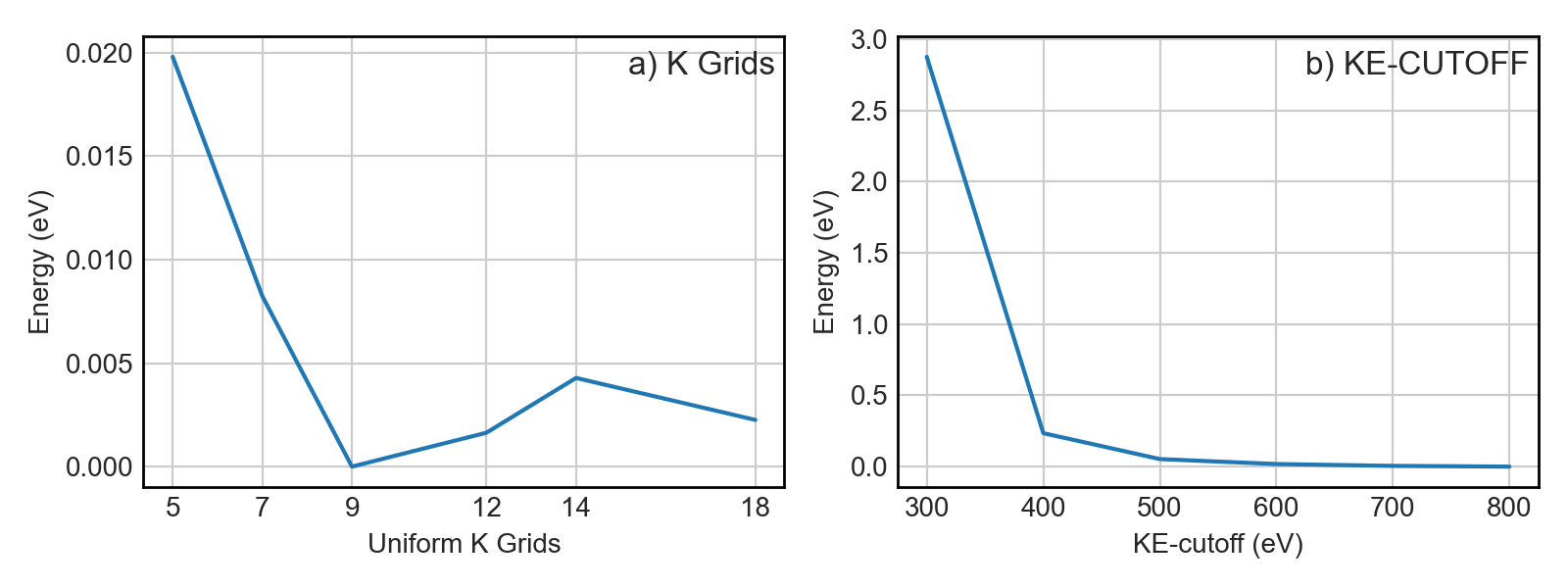}
    \caption{DFT Convergence test of P6/mmm: a) K-points test. KE-cutoff 400eV with XC functional PBE. b) KE cut-off test with K-points grids $14 \times 14  \times 14$ and XC functional PBE.}
    \label{fig:KE_K_convergence_P6mmm}
\end{figure}

\begin{figure}[th!]
    % \centering
    \includegraphics[width=1.\linewidth]{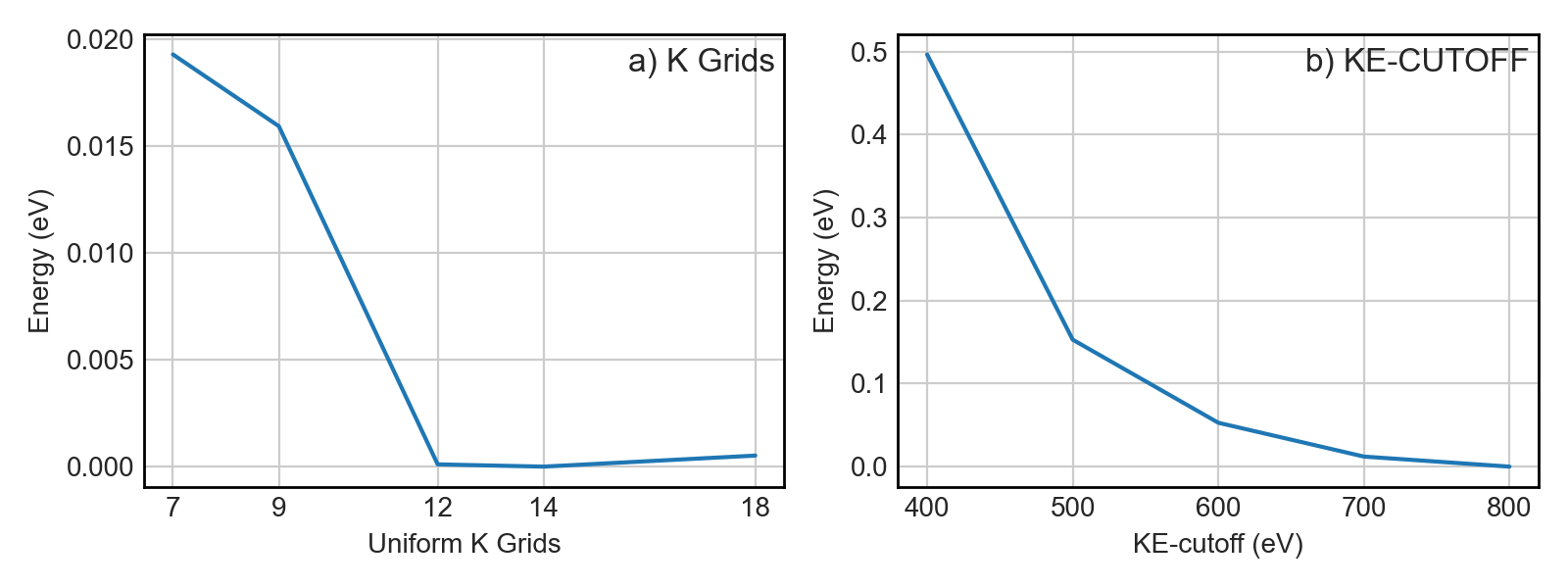}
    \caption{DFT Convergence test of $Im\overline{3}m$: a) K-points test. KE-cutoff 400eV with XC functional PBE. b) KE cut-off test with K-points grids $14 \times 14  \times 14$ and XC functional PBE.}
    \label{fig:KE_K_convergence_Im3m}
\end{figure}

\begin{figure}[th!]
    % \centering
    \includegraphics[width=1.\linewidth]{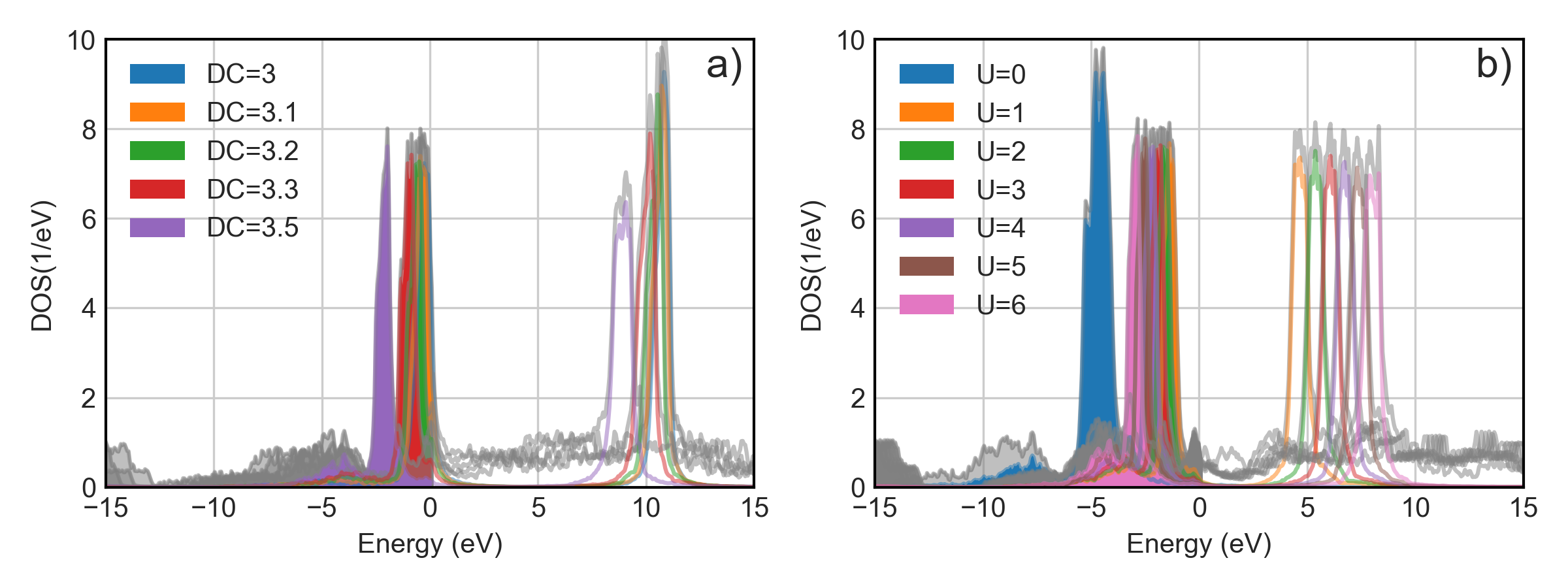}
    \caption{(a) DC test with DFT+DMFT CSC DOS for $P6/mmm$ SmH$_2$: Hubbard U equal to 6eV and Hund's coupling J equal to 0.855eV (b) Hubbard U test with 1-shot DOS for $P6/mmm$ SmH$_2$: DC equal 3.5 and Hund's coupling J equal 0.855. Notice that Converged DC for $P6/mmm$ is around 3.1}
    \label{fig:U-test}
\end{figure}

\end{document}